\documentclass[prd,twocolumn,showpacs,floatfix,amsmath,amssymb,floatfix,nofootinbib]{revtex4}
\usepackage{graphicx,color,dcolumn,booktabs,bm}
\usepackage{longtable,lscape,comment,braket}
\usepackage{txfonts}
\usepackage{overpic}
\usepackage{amssymb}
\usepackage{indentfirst}
\usepackage{feynmf}   
\usepackage{slashed}  
\usepackage{cases}
\usepackage{epstopdf}
\usepackage{psfrag}
\usepackage{subfigure}
\usepackage{threeparttable}
\pdfoutput=1
\usepackage{color}
\usepackage{multirow}

\graphicspath{{Figures/}} %
\usepackage[colorlinks, citecolor=blue,anchorcolor=red,menucolor=red, linkcolor=red,filecolor=red,runcolor=red,urlcolor=blue,frenchlinks=red]{hyperref}

\usepackage{ulem}
\def\be{\begin{equation}}
\def\ee{\end{equation}}

\begin{document}

\title{Prediction of anomalous $\Upsilon(5S)\to\Upsilon(1^3D_J)\eta$ transitions}
\author{Bo Wang}\email{wangb13@lzu.edu.cn}
\author{Xiang Liu}\email{xiangliu@lzu.edu.cn}
\affiliation{
Research Center for Hadron and CSR Physics, Lanzhou University and Institute of Modern Physics of CAS, Lanzhou 730000, China\\
School of Physical Science and Technology, Lanzhou University, Lanzhou 730000, China}

\author{Dian-Yong Chen}\email{chendy@seu.edu.cn}
\affiliation{Department of Physics, Southeast University, Nanjing 210094, China}

\begin{abstract}
In this work, we study the hadronic loop contribution to the $\Upsilon(5S)\to \Upsilon(1^3D_J)\eta$ ($J=1,2,3$) transitions. We predict that the branching ratios of $\Upsilon(5S)\to \Upsilon(1^3D_1)\eta$, $\Upsilon(5S)\to \Upsilon(1^3D_2)\eta$ and $\Upsilon(5S)\to \Upsilon(1^3D_3)\eta$ can reach up to $(0.5\sim5.1)\times10^{-3}$, $(0.7\sim7.5)\times10^{-3}$ and $(0.9\sim9.6)\times10^{-4}$, respectively. Since these predicted hadronic transitions of $\Upsilon(5S)$ are comparable with these observed $\Upsilon(5S)\to \Upsilon(nS)\pi^+\pi^-$ $(n=1,2,3)$, we suggest future experiment like Belle and BelleII to carry out the search for these anomalous $\Upsilon(5S)\to \Upsilon(1^3D_J)\eta$ transitions.

\end{abstract}

 \maketitle

\section{Introduction}\label{sec1}
Since 2007, the Belle Collaboration has focused on the hadronic transitions of higher bottomonium $\Upsilon(5S)$ experimentally. With the experimental progresses, more and more puzzling phenomena have been reported, which include anomalous $\Upsilon(5S)\to \Upsilon(nS)\pi^+\pi^-$ $(n=1,2,3)$ \cite{Abe:2007tk}, $\Upsilon(5S)\to \chi_{bJ}\omega$ \cite{He:2014sqj}, and the observed bottomonium-like states $Z_b(10610)/Z_b(10650)$ in the $\Upsilon(5S)\to \Upsilon(nS)\pi^+\pi^-$ and $\Upsilon(5S)\to h_b(mP)\pi^+\pi^-$ ($m=1,2$) processes \cite{Belle:2011aa}.

The former experience of studying these hadronic transitions of the $\Upsilon(5S)$ shows that the hadronic loop effect plays important role to solve these puzzles \cite{Meng:2007tk,Meng:2008dd,Simonov:2008qy,Chen:2011qx,Chen:2011zv,Chen:2011pv,Meng:2008bq,Chen:2014ccr}. In Ref. \cite{Meng:2007tk}, the hadronic loop mechanism was proposed to explain the large branching ratio of $\Upsilon(5S)\to \Upsilon(nS)\pi^+\pi^-$. In addition, the energy distribution of $\Upsilon(5S)\to \Upsilon(nS)\pi^+\pi^-$ was studied under this framework \cite{Meng:2008dd}. Later, by chiral Lagrangian approach, Simonov {\it et al.} \cite{Simonov:2008qy} also confirmed the conclusion made in Ref. \cite{Meng:2007tk}.  Chen {\it et al.} \cite{Chen:2011qx} further developed the hadronic loop mechanism by considering the interference between the direct hidden-charm dipion decay
and the transition via the intermediate hadronic loop, by which the anomalously large production rates and the lineshapes of the differential width of $\Upsilon(5S)\to \Upsilon(1S)\pi^+\pi^-$ can be reproduced. However, their study indicates a new puzzle, i.e., the obtained differential width $d\Gamma(\Upsilon(5S)\to \Upsilon(2S)\pi^+\pi^-)/d\cos\theta$ contradict with the experimental data while other results are well in accord with the data \cite{Meng:2007tk}. In Ref. \cite{Chen:2011zv}, Chen and Liu found that introducing the $Z_b(10610)/Z_b(10650)$ contribution can naturally explain this new puzzle existing in the $\Upsilon(5S)\to \Upsilon(2S)\pi^+\pi^-$ transition, which also stimulated the proposal of initial single pion emission mechanism in the hidden-bottom dipion decays of $\Upsilon(5S)$ \cite{Chen:2011pv}.
Besides these studies of the hidden-bottom dipion decays of the $\Upsilon(5S)$, the hadronic loop mechanism has been applied to investigate $\Upsilon(5S)\to \Upsilon(1S)\eta$ \cite{Meng:2008bq} and to explain the observed $\Upsilon(5S)\to \chi_{bJ}\omega$ decays \cite{Chen:2014ccr}.  As an important non-perturbative behaviors of QCD, the coupled-channel effect becomes obvious especially for higher charmonium/bottomonium. Here, we need to emphasize that the hadronic loop mechanism is as an equivalent description of the coupled-channel effect \cite{Meng:2007tk,Meng:2008dd,Chen:2011qx,Chen:2011zv,Chen:2011pv,Meng:2008bq,Chen:2014ccr,Liu:2006dq,Liu:2009dr,Li:2013zcr}.

It is obvious that it is not the end of whole story of the transitions of the $\Upsilon(5S)$.
Considering the above research status of the $\Upsilon(5S)$ hadronic transitions, we have a reason to believe that the hadronic loop mechanism can contribute to the transitions $\Upsilon(5S)\to\Upsilon(1^3D_J)\eta$ ($J=1,2,3$), where the $\Upsilon(1^3D_J)$ denotes the $D$-wave bottomonia \cite{Eichten:1980mw,Godfrey:1985xj,Ebert:2002pp}.
In this work, we perform the realistic study of the $\Upsilon(5S)\to\Upsilon(1^3D_J)\eta$ transition by introducing the hadronic loop contribution composed of bottom mesons. And then, the branching ratios of $\Upsilon(5S)\to\Upsilon(1^3D_J)\eta$ are predicted, by which we can learn that there may exist anomalous behavior of the $\Upsilon(5S)\to\Upsilon(1^3D_J)\eta$ transitions. Experimental search for these predicted $\Upsilon(5S)\to\Upsilon(1^3D_J)\eta$ will be an intriguing research issue, especially for Belle and forthcoming BelleII.
Until now, the $\Upsilon(1^3D_{2})$ has been reported in experiments \cite{Bonvicini:2004yj,delAmoSanchez:2010kz}. However, its partners like $\Upsilon(1^3D_{1})$ and $\Upsilon(1^3D_{3})$ are still missing in experiment. Thus, as a key point of exploring $\Upsilon(5S)\to\Upsilon(1^3D_J)\eta$, how to identify the $\Upsilon(1^3D_{1})$ and the $\Upsilon(1^3D_{3})$ will be an important research topic. In the following sections, we will give more discussions of this point.

This paper is organized as follows. After introduction, we present the detailed deduction of $\Upsilon(5S)\to\Upsilon(1^3D_J)\eta$ by considering the hadronic loop contribution in Sec. \ref{sec2}. And then, the corresponding numerical results will be given in Sec. \ref{sec3}. The paper ends with the discussion and conclusion.

\section{Hadronic loop effect on the $\Upsilon(5S)\to\Upsilon(1^3D_J)\eta$ decay}\label{sec2}

\begin{figure}[hptb]
	\scalebox{0.87}{\includegraphics[width=\columnwidth]{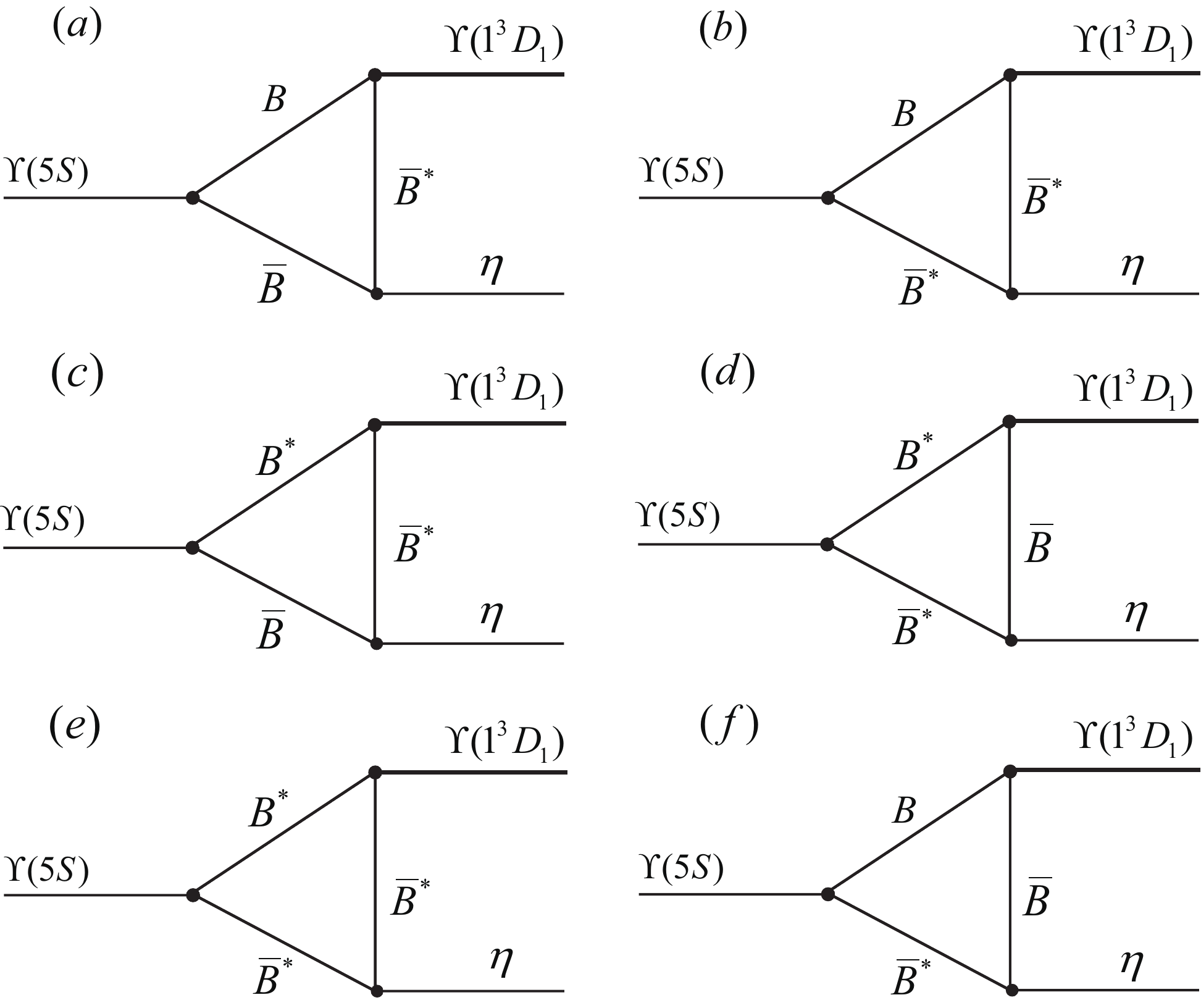}}
	\caption{The typical diagrams of the $\Upsilon(5S)$ decay into $\Upsilon(1^3D_1)\eta$ via the hadronic loop mechanism. Other similar diagrams composed of bottom-strange meson loop could be obtained by replacing the $B^{(\ast)}$ and $\bar{B}^{(\ast)}$ by $B_s^{(\ast) 0}$ and $\bar{B}_s^{(\ast) 0}$, respectively. \label{fig:1}}
\end{figure}

\begin{figure}[hptb]
	\scalebox{0.87}{\includegraphics[width=\columnwidth]{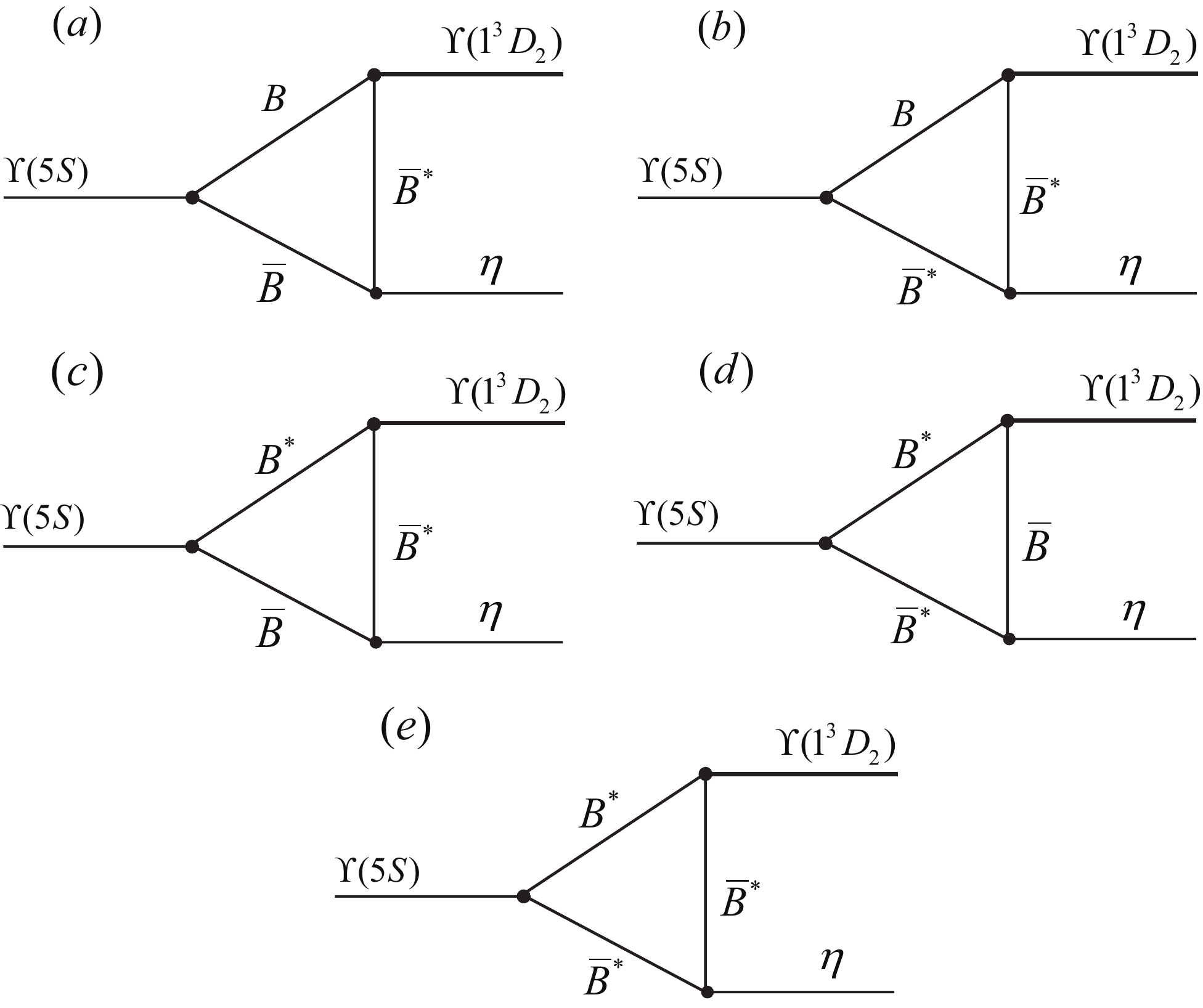}}
	\caption{The same as Fig. \ref{fig:1} but for $\Upsilon(5S) \to \Upsilon(1^3D_2) \eta$. \label{fig:3}}
\end{figure}

\begin{figure}[hptb]
	\scalebox{0.87}{\includegraphics[width=\columnwidth]{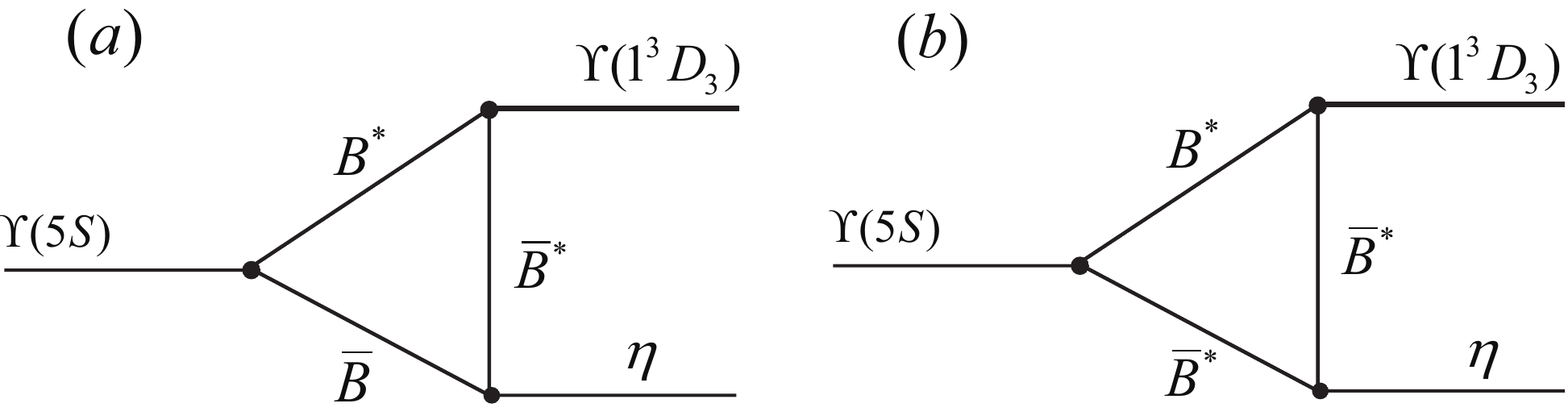}}
	\caption{The same as Fig. \ref{fig:1} but for $\Upsilon(5S) \to \Upsilon(1^3D_3) \eta$. \label{fig:4}}
\end{figure}

For calculating the hadronic transitions of $\Upsilon(5S)\to\Upsilon(1^3D_J)\eta$, we introduce the hadronic loop mechanism \cite{Meng:2007tk,Meng:2008dd,Chen:2011qx,Chen:2011zv,Chen:2011pv,Meng:2008bq,Chen:2014ccr,Liu:2006dq,Liu:2009dr,Li:2013zcr}. Under this mechanism, the transition $\Upsilon(5S)\to\Upsilon(1^3D_J)\eta$ occurs via the intermediate state $B_{(s)}^{(\ast)}\bar{B}_{(s)}^{(\ast)}$, which is as a bridge to connect the initial state $\Upsilon(5S)$ and final states $\Upsilon(1^3D_J)\eta$. The schematic diagrams for describing $\Upsilon(5S)\to\Upsilon(1^3D_J)\eta$ under the hadronic loop mechanism are listed in Figures \ref{fig:1}-\ref{fig:4}.

{In our calculations, we adopt the effective Lagrangian approach, where the relevant effective Lagrangians can be constructed by considering both heavy quark symmetry and chiral symmetry \cite{Casalbuoni:1996pg,Colangelo:2003sa}.
The pseudoscalar meson $B_{(s)}$ and the corresponding vector partner $B_{(s)}^\ast$ may form the spin doublet in the heavy quark limit $m_Q\to\infty$ \cite{Casalbuoni:1996pg}. This spin doublet can be represented by a $4\times4$ Dirac-type matrix $H$ \cite{Colangelo:2003sa,Casalbuoni:1996pg}, i.e.,
\begin{eqnarray}
H_{1}&=&\frac{1+\slashed{v}}{2}\left[B_{(s)}^{\ast\mu}\gamma_\mu-B_{(s)}\gamma_5\right] \label{h1},\\
H_{2}&=&\left[\bar{B}_{(s)}^{\ast\mu}\gamma_\mu-\bar{B}_{(s)}\gamma_5\right]\frac{1-\slashed{v}}{2}\label{h2},
\end{eqnarray}
where $H_1$ and $H_2$ represent the current with quark constituents $b\bar{q}$ and $\bar{b}q$, respectively, and have definition $\bar{H}_{1,2}=\gamma^0H_{1,2}^\dag\gamma^0$. A normalization factor $\sqrt{m_{B_{(s)}^{(\ast)}}}$ is contained in the $B_{(s)}^{(\ast)}$ field, where $m_{B_{(s)}^{(\ast)}}$ is the mass of $B_{(s)}^{(\ast)}$ meson.

Bottomonia with orbital angular momentum $L$ and different spin $S$ can be grouped into a multiplet. In our calculation, we consider the $S$-wave and $D$-wave bottomonia. For $S$-wave bottomonium doublet, it has the form \cite{Casalbuoni:1996pg},
\begin{eqnarray}
\mathcal{J}=\frac{1+\slashed{v}}{2}\Big[\Upsilon^\mu\gamma_\mu-\eta_b\gamma_5\Big]\frac{1-\slashed{v}}{2},\label{J0}
\end{eqnarray}
where $v^\mu$ is the 4-velocity of the heavy states, and $\Upsilon^\mu$ and $\eta_b$ correspond to $1^-$ and $0^-$ states, respectively. For $D$-wave bottomonium multiplet, we have
\begin{eqnarray}
\mathcal{J}^{\mu\lambda}&=&\frac{1+\slashed{v}}{2}\Bigg[\Upsilon_3^{\mu\lambda\alpha}\gamma_\alpha+\frac{1}{\sqrt{6}}\Big(\epsilon^{\mu\alpha\beta\rho}v_\alpha\gamma_\beta \Upsilon_{2\rho}^\lambda+\epsilon^{\lambda\alpha\beta\rho}v_\alpha\gamma_\beta \Upsilon_{2\rho}^\mu\Big)\nonumber\\
&&+\frac{\sqrt{15}}{10}\Big[(\gamma^\mu-v^\mu)\Upsilon_1^\lambda+(\gamma^\lambda-v^\lambda)\Upsilon_1^\mu\Big]\nonumber\\
&&-\frac{1}{\sqrt{15}}\Big(g^{\mu\lambda}-v^\mu v^\lambda\Big)\gamma_\alpha\Upsilon_1^\alpha+\eta_{b2}^{\mu\lambda}\gamma_5\Bigg]\frac{1-\slashed{v}}{2}, \label{J2}
\end{eqnarray}
where all the tensor fields in Eq. (\ref{J2}) are traceless, symmetric and transverse. The fields $\Upsilon_3$, $\Upsilon_2$, $\Upsilon_1$ and $\eta_{b2}$ denote the bottomonia with $J^{PC}=3^{--}$, $2^{--}$, $1^{--}$ and $2^{-+}$, respectively. The general effective Lagrangians describing the interactions of the open-bottom mesons with the $S$-wave or $D$-wave bottomonium multiplet are
\begin{eqnarray} \label{XHH}
\mathcal{L}_1 &=&g_1 \mathrm{Tr}\Big[\mathcal{J}\bar{H}_{2}\overleftrightarrow{\partial}_\mu\gamma^\mu\bar{H}_{1}\Big]+\mathrm{H.c.},\nonumber\\
\mathcal{L}_2 &=&g_2 \mathrm{Tr}\Big[\mathcal{J}^{\mu\lambda}\bar{H}_{2}\overleftrightarrow{\partial}_\mu\gamma_\lambda\bar{H}_{1}\Big]+\mathrm{H.c.} . \label{lag1}
\end{eqnarray}

The effective Lagrangian for the strong interactions between heavy mesons and light pseudoscalars can be established when considering both heavy quark $SU(2N_f)$ symmetry and chiral $SU(3)_L\bigotimes SU(3)_R$ symmetry. The octet of light pseudoscalar field is represented by the coset field $\xi(x)=e^{i\mathbf{\mathcal{M}}/f}$ with
\begin{eqnarray}
\mathbf{\mathcal{M}} =
\left( \begin{array}{ccc}
\sqrt{\frac{1}{2}}\pi^0+\sqrt{\frac{1}{6}}\eta_8 & \pi^+ & K^+ \\
\pi^- & -\sqrt{\frac{1}{2}}\pi^0+\sqrt{\frac{1}{6}}\eta_8 & K^0 \\
K^- & \bar{K}^0 & -\sqrt{\frac{2}{3}}\eta_8
\end{array} \right),
\end{eqnarray}
 and $f\simeq f_\pi=131$ MeV, and the effective Lagrangian can be written as
\begin{eqnarray}\label{lag3}
\mathcal{L}_3=ig_3 \mathrm{Tr}\Big[H_b\gamma^\mu\gamma^5\mathcal{A}_{ba}^\mu\bar{H}_a\Big],\label{lag7}
\end{eqnarray}
where the  axial current $\mathcal{A}^\mu$ is defined as $\mathcal{A}^\mu=1/2(\xi^\dag\partial_\mu\xi-\xi\partial_\mu\xi^\dag)$, and the leading order expansion of $\mathcal{A}_\mu$ gives $\mathcal{A}_\mu= i/f\partial_\mu\mathcal{M}$.
By expanding Eqs. (\ref{lag1}) and (\ref{lag7}), we can further obtain the specific effective Lagrangians involved in our calculations:
\begin{eqnarray}
\mathcal{L}_{\Upsilon \mathcal{B}^{(\ast)}\mathcal{B}^{(\ast)}}&=&g_{\Upsilon \mathcal{B}\mathcal{B}}\Upsilon_\mu(\mathcal{B}^\dag\partial^\mu \mathcal{B}-\mathcal{B}\partial^\mu \mathcal{B}^\dag),\label{eq1}\nonumber\\
&&+ig_{\Upsilon \mathcal{B}\mathcal{B}^\ast}\epsilon_{\alpha\rho\beta\lambda}\Big[\mathcal{B}^\dag\overleftrightarrow{\partial}^\alpha \mathcal{B}^{\ast\rho}-\mathcal{B}^{\ast\rho\dag}\overleftrightarrow{\partial}^\alpha \mathcal{B}\Big]{\partial}^\beta \Upsilon^\lambda,\nonumber\\
&&+g_{\Upsilon \mathcal{B}^\ast \mathcal{B}^\ast}\Big[-(\Upsilon^\mu \mathcal{B}^{\ast\nu}\partial_\mu \mathcal{B}^{\ast\nu\dag}-\Upsilon^\mu \mathcal{B}^{\ast\nu\dag}\partial_\mu \mathcal{B}^{\ast\nu})\nonumber\\
&&+\Upsilon^\mu \mathcal{B}^{\ast\nu}\partial_\nu \mathcal{B}^{\ast\dag}_\mu-\Upsilon^\mu \mathcal{B}^{\ast\nu\dag}\partial_\nu \mathcal{B}^{\ast}_\mu\Big],\\
\mathcal{L}_{\Upsilon_J \mathcal{B}^{(\ast)}\mathcal{B}^{(\ast)}}&=&g_{\Upsilon_1 \mathcal{B}\mathcal{B}}\Upsilon_1^\mu(\mathcal{B}^\dag\partial_\mu \mathcal{B}-\mathcal{B}\partial_\mu \mathcal{B}^\dag),\nonumber\\
&&+ig_{\Upsilon_1 \mathcal{B}\mathcal{B}^\ast}\epsilon^{\mu\nu\alpha\beta}\Big[\mathcal{B}^\dag\overleftrightarrow{\partial}_\mu \mathcal{B}_\beta^\ast-\mathcal{B}_\beta^{\ast\dag}\overleftrightarrow{\partial}_\mu \mathcal{B}\Big]\partial_\nu\Upsilon_1^\alpha,\nonumber\\
&&+g_{\Upsilon_1 \mathcal{B}^\ast \mathcal{B}^\ast}\Big[-4(\Upsilon_1^\mu \mathcal{B}^{\ast\nu}\partial_\mu \mathcal{B}^{\ast\nu\dag}-\Upsilon_1^\mu \mathcal{B}^{\ast\nu\dag}\partial_\mu \mathcal{B}^{\ast\nu})\nonumber\\
&&+\Upsilon_1^\mu \mathcal{B}^{\ast\nu}\partial_\nu \mathcal{B}^{\ast\dag}_\mu-\Upsilon_1^\mu \mathcal{B}^{\ast\nu\dag}\partial_\nu \mathcal{B}^{\ast}_\mu\Big],\nonumber\\
&&+ig_{\Upsilon_{2} \mathcal{B}\mathcal{B}^\ast}\Upsilon_{2}^{\mu\nu}(\mathcal{B}^\dag\overleftrightarrow{\partial}_\nu \mathcal{B}_\mu^\ast-\mathcal{B}_\mu^{\ast\dag}\overleftrightarrow{\partial}_\nu \mathcal{B}),\nonumber\\
&&+g_{\Upsilon_{2} \mathcal{B}^\ast \mathcal{B}^\ast}\epsilon^{\alpha\beta\mu\nu}\Big[\mathcal{B}^{\ast^\nu\dag}\overleftrightarrow{\partial}^\beta \mathcal{B}^{\ast\lambda}+\mathcal{B}^{\ast^\nu}\overleftrightarrow{\partial}^\beta \mathcal{B}^{\ast\lambda\dag}\Big]\partial_\mu\Upsilon_{2}^{\alpha\lambda},\nonumber\\
&&+g_{\Upsilon_3\mathcal{B}^\ast \mathcal{B}^\ast}\Upsilon_3^{\mu\nu\alpha}\Big[\mathcal{B}_\alpha^{\ast\dag}\overleftrightarrow{\partial}_\mu \mathcal{B}_\nu^\ast+\mathcal{B}_\nu^{\ast\dag}\overleftrightarrow{\partial}_\mu \mathcal{B}_\alpha^\ast\Big],\\
\mathcal{L}_{\eta\mathcal{B}^{(\ast)}\mathcal{B}^{(\ast)}}&=&g_{\mathcal{B}^\ast \mathcal{B}\eta}[\mathcal{B}_\mu^\ast\partial^\mu\eta \mathcal{B}^\dag+\mathcal{B}_\mu^{\ast\dag}\partial^\mu \eta \mathcal{B}],\nonumber\\
&&+ig_{\mathcal{B}^\ast \mathcal{B}^\ast\eta}\epsilon^{\mu\nu\alpha\beta}\partial_\mu \eta\partial_\nu \mathcal{B}_\alpha^{\ast} \mathcal{B}_\beta^{\ast\dag}.\label{eq11}
\end{eqnarray}
In the above expressions, $\Upsilon$ and $\Upsilon_J$ are the abbreviations of the $\Upsilon(5S)$ and the $\Upsilon(1^3D_J)$, respectively, and $\mathcal{B}^{(\ast)\dag}$ and $\mathcal{B}^{(\ast)}$ have the definitions $\mathcal{B}^{(\ast)\dag}=(B^{(\ast)+},B^{(\ast)0},B_s^{(\ast)0})$ and $\mathcal{B}^{(\ast)}=(B^{(\ast)-},\bar{B}^{(\ast)0},\bar{B}_s^{(\ast)0})^T$. Later, we will list the concrete values of the coupling constants appearing in Eqs. (\ref{eq1})-(\ref{eq11}).

With the above preparation, we write out the amplitudes of Fig. \ref{fig:1}, which corresponds to $\Upsilon(5S)\to B^{(\ast)}(p_1)\bar{B}^{(\ast)}(p_2)\to\Upsilon(1^3D_1)(k_1)\eta(k_2)$, i.e.,
\begin{eqnarray}\label{amp1}
\mathcal{A}_{(a)}^{(1)}&=&(i)^3\int\frac{d^4q}{(2\pi)^4}\Big[g_{\Upsilon BB}(ip_2^\mu-ip_1^\mu)\epsilon_\Upsilon^\mu\Big]\nonumber\\
&&\times\Big[ig_{\Upsilon_{1}BB^\ast}\epsilon^{\theta\eta\alpha\beta}(-iq_\theta+ip_{1\theta})ik_{1\eta}\epsilon_{\Upsilon_{1}}^{\ast\alpha}\Big]\Big[g_{BB^\ast\eta}ik_2^\rho\Big]\nonumber\\
&&\times\frac{1}{p_1^2-m_B^2}\frac{1}{p_2^2-m_B^2}\frac{-g^{\beta\rho}+q^\beta q^\rho/m_{B^\ast}^2}{q^2-m_{B^\ast}^2}\mathcal{F}^2(q^2),
\end{eqnarray}
\begin{eqnarray}
\mathcal{A}_{(b)}^{(1)}&=&(i)^3\int\frac{d^4q}{(2\pi)^4}\Big[ig_{\Upsilon BB^\ast}\epsilon_{\alpha\rho\beta\lambda}(ip_4^\alpha-ip_3^\alpha)(-ip^\beta)\epsilon_\Upsilon^\lambda \Big]\nonumber\\
&&\times\Big[ig_{\Upsilon_{1}BB^\ast}\epsilon^{\kappa\sigma\tau\omega}(-iq_\kappa+ip_{3\kappa})ik_{1\sigma}\epsilon_{\Upsilon_{1}}^{\ast\tau}\Big]\nonumber\\
&&\times\Big[ig_{B^\ast B^\ast\eta}\epsilon^{\gamma\delta\theta\eta}(i q_{\delta})ik_{2\gamma}\Big]\frac{1}{p_3^2-m_B^2}\nonumber\\
&&\times\frac{-g^{\rho\eta}+p_4^\rho p_4^\eta/m_{B^\ast}^2}{p_4^2-m_{B^\ast}^2}\frac{-g^{\omega\theta}+q^\omega q^\theta/m_{B^\ast}^2}{q^2-m_{B^\ast}^2}\mathcal{F}^2(q^2),
\end{eqnarray}
\begin{eqnarray}
\mathcal{A}_{(c)}^{(1)}&=&(i)^3\int\frac{d^4q}{(2\pi)^4}\Big[ig_{\Upsilon BB^\ast}\epsilon^{\alpha\rho\beta\lambda}(ip_4^\alpha-ip_3^\alpha)(-ip^\beta)\epsilon_\Upsilon^\lambda\Big]\nonumber\\
&&\times\Big[g_{\Upsilon_1B^\ast B^\ast}\epsilon_{\Upsilon_1}^{\ast\mu}(4(iq_\mu-ip_{4\mu})g_{\nu\delta}+(ip_{4\nu}-iq_\nu)g_{\mu\delta})\Big]\nonumber\\
&&\times\Big[g_{BB^\ast\eta}ik_2^\eta\Big]\frac{1}{p_3^2-m_B^2}\frac{-g^{\nu\rho}+p_4^\nu p_4^\rho/m_{B^\ast}^2}{p_4^2-m_{B^\ast}^2}\nonumber\\
&&\times\frac{-g^{\eta\delta}+q^\eta q^\delta/m_{B^\ast}^2}{q^2-m_{B^\ast}^2}\mathcal{F}^2(q^2),
\end{eqnarray}
\begin{eqnarray}
\mathcal{A}_{(d)}^{(1)}&=&(i)^3\int\frac{d^4q}{(2\pi)^4}\Big[g_{\Upsilon B^\ast B^\ast}\epsilon_\Upsilon^\mu((ip_6^\mu-ip_5^\mu)g_{\nu\lambda}\nonumber\\
&&+(ip_5^\nu-ip_6^\nu)g_{\mu\lambda})\Big]\Big[ig_{\Upsilon_{1}BB^\ast}\epsilon^{\theta\eta\alpha\beta}(-ip_5^\theta+ iq^\theta)ik_{1\eta}\epsilon_{\Upsilon_1}^{\ast\alpha}\Big]\nonumber\\
&&\times\Big[g_{BB^\ast\eta}ik_2^\rho\Big]\frac{1}{q^2-m_B^2}\frac{-g^{\beta\nu}+p_5^\beta p_5^\nu/m_{B^\ast}^2}{p_5^2-m_{B^\ast}^2}\nonumber\\
&&\times\frac{-g^{\rho\lambda}+p_6^\rho p_6^\lambda/m_{B^\ast}^2}{p_6^2-m_{B^\ast}^2}\mathcal{F}^2(q^2),
\end{eqnarray}
\begin{eqnarray}
\mathcal{A}_{(e)}^{(1)}&=&(i)^3\int\frac{d^4q}{(2\pi)^4}\Big[g_{\Upsilon B^\ast B^\ast}\epsilon_\Upsilon^\delta((ip_6^\delta-ip_5^\delta)g_{\phi\lambda}\nonumber\\
&&+(ip_5^\phi-ip_6^\phi)g_{\delta\lambda})\Big]\times\Big[g_{\Upsilon_{1}B^\ast B^\ast}(4(iq^\mu-ip_5^\mu)g_{\nu\alpha}\nonumber\\
&&+(ip_{5\nu}-iq_{\nu})g_{\mu\alpha})\epsilon^{\ast\mu}_{\Upsilon_1}\Big]\Big[ig_{B^\ast B^\ast\eta}\epsilon^{\gamma\sigma\kappa\eta}(iq^\sigma)ik_2^\gamma\Big]\nonumber\\
&&\times\frac{-g^{\phi\nu}+p_5^\phi p_5^\nu/m_{B^\ast}^2}{p_5^2-m_{B^\ast}^2}\frac{-g^{\eta\lambda}+p_6^\eta p_6^\lambda/m_{B^\ast}^2}{p_6^2-m_{B^\ast}^2}\nonumber\\
&&\times\frac{-g^{\kappa\alpha}+q^\kappa q^\alpha/m_{B^\ast}^2}{q^2-m_{B^\ast}^2}\mathcal{F}^2(q^2),
\end{eqnarray}
\begin{eqnarray}
\mathcal{A}_{(f)}^{(1)}&=&(i)^3\int\frac{d^4q}{(2\pi)^4}\Big[ig_{\Upsilon BB^\ast}\epsilon_{\alpha\rho\beta\lambda}(ip_4^\alpha-ip_3^\alpha)(-ip^\beta)\epsilon_\Upsilon^\lambda \Big]\nonumber\\
&&\times\Big[g_{\Upsilon_{1}BB}(iq_\mu-ip_{3\mu})\epsilon_{\Upsilon_{1}}^{\ast\mu}\Big]\Big[g_{B B^\ast\eta}ik_{2\nu}\Big]\nonumber\\
&&\times\frac{1}{p_3^2-m_B^2}\frac{-g^{\rho\nu}+p_4^\rho p_4^\nu/m_{B^\ast}^2}{p_4^2-m_{B^\ast}^2}\frac{1}{q^2-m_{B}^2}\mathcal{F}^2(q^2),
\end{eqnarray}
where the monopole form factor $\mathcal{F}(q^2)$ included in above amplitudes is for depicting the off-shell effect of the exchanged meson of the rescattering process $B^{(\ast)}\bar{B}^{(\ast)} \to\Upsilon(1^3D_1)\eta$, which can be represented as \cite{Colangelo:2003sa,Cheng:2004ru}
\begin{eqnarray}
\label{ff}
\mathcal{F}(q^2)=\frac{m_{E}^2-\Lambda^2}{q^2-\Lambda^2},\ \  \Lambda=m_E+\alpha\Lambda_{QCD},
\end{eqnarray}
where $m_E$ and $q$ denote the mass and four-momentum of the exchanged meson, respectively. $\Lambda_{QCD}=220$ MeV and $\alpha$ is a phenomenological dimensionless parameter.

The total transition amplitude for $\Upsilon(5S)\to\Upsilon(1^3D_1)\eta$ can be further expressed as
\begin{eqnarray}\label{sumamp2}
\mathcal{A}[\Upsilon(5S)\to\Upsilon(1^3D_1)\eta]=4\sum_{i=a,...,f}\mathcal{A}^{(1)}_{(i) }+2\sum_{i=a,...,f}\tilde{\mathcal{A}}^{(1)}_{(i)},\label{rh}
\end{eqnarray}
where the factor 4 in the first term on the right hand side of Eq. (\ref{rh}) comes from the charge conjugation transformation $B^{(\ast)}\leftrightarrow\bar{B}^{(\ast)}$ and isospin transformations $B^{(\ast)0}\leftrightarrow B^{(\ast)+}$ and $\bar{B}^{(\ast)0}\leftrightarrow B^{(\ast)-}$. While the factor 2 in the second term is due to the charge conjugation transformation $B_s^{(\ast)}\leftrightarrow \bar{B}_s^{(\ast)}$. Here, $\tilde{\mathcal{A}}_{(i)}$ denotes that the amplitude contribution is from intermediate $B_s^{(\ast)}\bar{B}_s^{(\ast)}$ state.

For the $\Upsilon(5S)\to\Upsilon(1^3D_{2})\eta$ decay, the amplitudes corresponding to Fig. \ref{fig:3} (a)-(e) read
\begin{eqnarray}\label{absa}
\mathcal{A}^{(2)}_{(a)}&=&(i)^3\int\frac{d^4q}{(2\pi)^4}\Big[g_{\Upsilon BB}(ip_2^\mu-ip_1^\mu)\epsilon_\Upsilon^\mu\Big]\nonumber\\
&&\times\Big[ig_{\Upsilon_{2}BB^\ast}(-iq_\beta+ip_{1\beta})\epsilon_{\Upsilon_{2}}^{\ast\alpha\beta}\Big]\Big[g_{BB^\ast\eta}ik_2^\rho\Big]\nonumber\\
&&\times\frac{1}{p_1^2-m_B^2}\frac{1}{p_2^2-m_B^2}\frac{-g^{\alpha\rho}+q^\alpha q^\rho/m_{B^\ast}^2}{q^2-m_{B^\ast}^2}\mathcal{F}^2(q^2),
\end{eqnarray}
\begin{eqnarray}
\mathcal{A}^{(2)}_{(b)}&=&(i)^3\int\frac{d^4q}{(2\pi)^4}\Big[ig_{\Upsilon BB^\ast}\epsilon_{\alpha\rho\beta\lambda}(ip_4^\alpha-ip_3^\alpha)(-ip^\beta)\epsilon_\Upsilon^\lambda \Big]\nonumber\\
&&\times\Big[ig_{\Upsilon_{2} BB^\ast}(-iq_\nu+ip_{3\nu})\epsilon_{\Upsilon_{2}}^{\ast\mu\nu}\Big]\Big[ig_{B^\ast B^\ast\eta}\epsilon^{\gamma\delta\theta\eta}(i q_{\delta})ik_{2\gamma}\Big]\nonumber\\
&&\times\frac{1}{p_3^2-m_B^2}\frac{-g^{\rho\eta}+p_4^\rho p_4^\eta/m_{B^\ast}^2}{p_4^2-m_{B^\ast}^2}\nonumber\\
&&\times\frac{-g^{\mu\theta}+q^\mu q^\theta/m_{B^\ast}^2}{q^2-m_{B^\ast}^2}\mathcal{F}^2(q^2),
\end{eqnarray}
\begin{eqnarray}
\mathcal{A}^{(2)}_{(c)}&=&(i)^3\int\frac{d^4q}{(2\pi)^4}\Big[ig_{\Upsilon BB^\ast}\epsilon^{\alpha\rho\beta\lambda}(ip_4^\alpha-ip_3^\alpha)(-ip^\beta)\epsilon_\Upsilon^\lambda\Big]\nonumber\\
&&\times\Big[g_{\Upsilon_{2}B^\ast B^\ast}\epsilon^{\gamma\delta\mu\nu}(ip_4^\delta-iq^\delta)ik_1^\mu\epsilon_{\Upsilon_{2}}^{\ast \gamma\theta}\Big]\Big[g_{BB^\ast\eta}ik_2^\eta\Big]\nonumber\\
&&\times\frac{1}{p_3^2-m_B^2}\frac{-g^{\nu\rho}+p_4^\nu p_4^\rho/m_{B^\ast}^2}{p_4^2-m_{B^\ast}^2}\nonumber\\
&&\times\frac{-g^{\eta\theta}+q^\eta q^\theta/m_{B^\ast}^2}{q^2-m_{B^\ast}^2}\mathcal{F}^2(q^2),
\end{eqnarray}
\begin{eqnarray}
\mathcal{A}^{(2)}_{(d)}&=&(i)^3\int\frac{d^4q}{(2\pi)^4}\Big[g_{\Upsilon B^\ast B^\ast}\epsilon_\Upsilon^\mu((ip_6^\mu-ip_5^\mu)g_{\nu\lambda}\nonumber\\
&&+(ip_5^\nu-ip_6^\nu)g_{\mu\lambda})\Big]\Big[ig_{\Upsilon_{2}BB^\ast}\epsilon_{\Upsilon_{2}}^{\ast\alpha\beta}(-ip_5^\beta+ iq^\beta)\Big]\nonumber\\
&&\times\Big[g_{BB^\ast\eta}ik_2^\rho\Big]\frac{1}{q^2-m_B^2}\frac{-g^{\alpha\nu}+p_5^\alpha p_5^\nu/m_{B^\ast}^2}{p_5^2-m_{B^\ast}^2}\nonumber\\
&&\times\frac{-g^{\rho\lambda}+p_6^\rho p_6^\lambda/m_{B^\ast}^2}{p_6^2-m_{B^\ast}^2}\mathcal{F}^2(q^2),
\end{eqnarray}
\begin{eqnarray}
\mathcal{A}^{(2)}_{(e)}&=&(i)^3\int\frac{d^4q}{(2\pi)^4}\Big[g_{\Upsilon B^\ast B^\ast}\epsilon_\Upsilon^\delta((ip_6^\delta-ip_5^\delta)g_{\phi\lambda}\nonumber\\
&&+(ip_5^\phi-ip_6^\phi)g_{\delta\lambda})\Big]\Big[g_{\Upsilon_{2}B^\ast B^\ast}\epsilon^{\alpha\beta\mu\nu}(-ip_5^\beta+iq^\beta)ik_1^\mu\epsilon_{\Upsilon_{2}}^{\ast\alpha\tau}\Big]\nonumber\\
&&\times\Big[ig_{B^\ast B^\ast\eta}\epsilon^{\gamma\sigma\kappa\eta}(iq^\sigma)ik_2^\gamma\Big]\frac{-g^{\phi\tau}+p_5^\phi p_5^\tau/m_{B^\ast}^2}{p_5^2-m_{B^\ast}^2}\nonumber\\
&&\times\frac{-g^{\eta\lambda}+p_6^\eta p_6^\lambda/m_{B^\ast}^2}{p_6^2-m_{B^\ast}^2}\frac{-g^{\kappa\nu}+q^\kappa q^\nu/m_{B^\ast}^2}{q^2-m_{B^\ast}^2}\mathcal{F}^2(q^2).
\end{eqnarray}

For $\Upsilon(5S)\to\Upsilon(1^3D_{3})\eta$, the amplitudes relevant to Fig. \ref{fig:4} (a)-(b) can be written as
\begin{eqnarray}\label{amp2}
\mathcal{A}_{(a)}^{(3)}&=&(i)^3\int\frac{d^4q}{(2\pi)^4}\Big[ig_{\Upsilon BB^\ast}\epsilon^{\alpha\rho\beta\lambda}(ip_4^\alpha-ip_3^\alpha)(-ip^\beta)\epsilon_\Upsilon^\lambda\Big]\nonumber\\
&&\times\Big[g_{\Upsilon_3B^\ast B^\ast}\epsilon_{\Upsilon_3}^{\ast\mu\nu\delta}(iq_\mu-ip_{4\mu})\Big]\nonumber\\
&&\times\Big[g_{BB^\ast\eta}ik_2^\eta\Big]\frac{1}{p_3^2-m_B^2}\frac{-g^{\nu\rho}+p_4^\nu p_4^\rho/m_{B^\ast}^2}{p_4^2-m_{B^\ast}^2}\nonumber\\
&&\times\frac{-g^{\eta\delta}+q^\eta q^\delta/m_{B^\ast}^2}{q^2-m_{B^\ast}^2}\mathcal{F}^2(q^2),
\end{eqnarray}
\begin{eqnarray}
\mathcal{A}_{(b)}^{(3)}&=&(i)^3\int\frac{d^4q}{(2\pi)^4}\Big[g_{\Upsilon B^\ast B^\ast}\epsilon_\Upsilon^\delta((ip_6^\delta-ip_5^\delta)g_{\phi\lambda}\nonumber\\
&&+(ip_5^\phi-ip_6^\phi)g_{\delta\lambda})\Big]\Big[g_{\Upsilon_3B^\ast B^\ast}\epsilon_{\Upsilon_3}^{\ast\mu\nu\delta}(iq_\mu-ip_{5\mu})\Big]\nonumber\\
&&\times\Big[ig_{B^\ast B^\ast\eta}\epsilon^{\gamma\sigma\kappa\eta}(iq^\sigma)ik_2^\gamma\Big]\frac{-g^{\phi\nu}+p_5^\phi p_5^\nu/m_{B^\ast}^2}{p_5^2-m_{B^\ast}^2}\nonumber\\
&&\times\frac{-g^{\eta\lambda}+p_6^\eta p_6^\lambda/m_{B^\ast}^2}{p_6^2-m_{B^\ast}^2}\frac{-g^{\kappa\delta}+q^\kappa q^\delta/m_{B^\ast}^2}{q^2-m_{B^\ast}^2}\mathcal{F}^2(q^2).
\end{eqnarray}

Therefore, the total amplitudes of $\Upsilon(5S)\to\Upsilon(1^3D_2)\eta$ and $\Upsilon(5S)\to\Upsilon(1^3D_3)\eta$ can be represented as
\begin{eqnarray}\label{sumamp}
\mathcal{A}[\Upsilon(5S)\to\Upsilon(1^3D_2)\eta]&=&4\sum_{i=a,...,e}\mathcal{A}_{(i)}^{(2)}+2\sum_{i=a,...,e}\mathcal{\tilde A}_{(i)}^{(2)},\nonumber\\
\mathcal{A}[\Upsilon(5S)\to\Upsilon(1^3D_3)\eta]&=&4\sum_{i=a,b}\mathcal{A}_{(i)}^{(3)}+2\sum_{i=a,b}\mathcal{\tilde A}_{(i)}^{(3)},\nonumber
\end{eqnarray}
where $\mathcal{\tilde A}_{(i)}^{(2)}$ and $\mathcal{\tilde A}_{(i)}^{(3)}$ denote the amplitudes
from $\Upsilon(5S)\to\Upsilon(1^3D_2)\eta$ and $\Upsilon(5S)\to\Upsilon(1^3D_3)\eta$, respectively, via intermediate $B_s^{(\ast)}\bar{B}_s^{(\ast)}$.


The coupling constants involved in Eqs. (\ref{eq1})-(\ref{eq11}) can be determined from the experimental data and some relations based on heavy quark limit and chiral symmetry. The coupling constants of the interactions of $\Upsilon(5S)$ with $B_{(s)}^{(*)}\bar{B}_{(s)}^{(*)}$ can be fixed by the partial widths of $\Upsilon(5S)\to B_{(s)}^{(*)}\bar B_{(s)}^{(*)}$ \cite{pdg2014}.
Under heavy quark symmetry, $g_{\Upsilon_J B^{(\ast)} B^{(\ast)}}$ can be expressed as
 \begin{subequations}
 \begin{eqnarray}\label{coups1}
g_{\Upsilon_1 B B}&=&-2g_2\frac{\sqrt{15}}{3}\sqrt{m_{\Upsilon_1}m_{B}m_{B}},\label{hhh1}\\
g_{\Upsilon_1 BB^\ast}&=&g_2\frac{\sqrt{15}}{3}\sqrt{m_B m_{B^\ast}/m_{\Upsilon_1}},\label{hhh3}\\
g_{\Upsilon_1 B^\ast B^\ast}&=&g_2\frac{\sqrt{15}}{15}\sqrt{m_{\Upsilon_1}m_{B^\ast}m_{B^\ast}},\\
g_{\Upsilon_{2} BB^\ast}&=&2g_2\sqrt{\frac{3}{2}}\sqrt{m_{\Upsilon_{2}}m_B m_{B^\ast}},\label{eq15}\\
g_{\Upsilon_{2} B^\ast B^\ast}&=&-2g_2\sqrt{\frac{1}{6}}\sqrt{m_{B^\ast} m_{B^\ast}/m_{\Upsilon_{2}}},\label{eq16}\\
g_{\Upsilon_3B^\ast B^\ast}&=&2g_2\sqrt{m_{\Upsilon_3}m_{B^\ast}m_{B^\ast}},\label{hhh2}
\end{eqnarray}
\end{subequations}
where $g_2$ can be fixed by the following approach. We first consider $\Upsilon(1^3D_1)$ state, which is a vector meson. Thus, its decay constant $f_\mathcal{V}$ can be defined as \cite{Neubert:1992}
\begin{eqnarray}
\langle0|Q\gamma^\mu\bar{Q}|\mathcal{V}\rangle=f_\mathcal{V}M_\mathcal{V}\epsilon^\mu_\mathcal{V},
\end{eqnarray}
where $M_\mathcal{V}$ and $\epsilon^\mu_\mathcal{V}$
denote the mass and polarization vector of vector meson, respectively. The decay constant $f_\mathcal{V}$ is related to the electromagnetic decay \cite{Neubert:1992}, i.e., \begin{eqnarray}
\Gamma_{\mathcal{V}\to e^+e^-}=\frac{4\pi}{3}\frac{\alpha^2}{M_\mathcal{V}}f_\mathcal{V}^2C_\mathcal{V},
\end{eqnarray}
where $\alpha$ is the fine-structure constant and $C_\mathcal{V}=1/9$ for $\Upsilon(1^3D_1)$ meson.

 In Ref. \cite{Godfrey:2015dia}, Godfrey {\it et al.} have calculated the the leptonic decay width of $\Upsilon(1^3D_1)$ in the framework of semi-relativistic quark model, i.e., $\Gamma[\Upsilon(1^3D_1)\to e^+e^-]=1.38$ eV. Thus, we can get decay constant $f_{\Upsilon_1}=23.8$ MeV. By the relation $g_{\Upsilon_1BB}\simeq M_{\Upsilon_1}/f_{\Upsilon_1}$ under vector meson dominance ansatz \cite{Colangelo:2003sa,Lin:1999ad,Deandrea:2003pv} and Eq. (\ref{hhh1}), then we obtain $g_2=9.83$ GeV$^{-3/2}$, which can be applied to estimate other coupling constants listed in Eqs. (\ref{hhh3})-(\ref{hhh2}), as shown in Table \ref{tab:1}.

In the $SU(3)$ quark model, $\eta$ and $\eta^\prime$ are mixings between octet $\eta_8$ and singlet $\eta_0$,
\begin{eqnarray}\label{mixing}
|\eta\rangle=\cos\theta|\eta_8\rangle-\sin\theta|\eta_0\rangle,\ \ \ |\eta^\prime\rangle=\sin\theta|\eta_8\rangle+\cos\theta|\eta_0\rangle,
\end{eqnarray}
where the mixing angle $\theta=-19.1^\circ$ is fixed by the experimental data \cite{Coffman:1988ve,Jousset:1988ni}. Additionally, there exist the relations of the coupling constants, i.e.,
\begin{eqnarray}
g_{BB^\ast\eta}&=&-\frac{1}{2}g_{B_sB_s^\ast\eta}=\frac{\sqrt{2}}{\sqrt{3}}\frac{\sqrt{m_Bm_{B^\ast}}}{f_\pi}g_3,\label{ff1}\\
g_{B^\ast B^\ast\eta}&=&\frac{g_{BB^\ast\eta}}{m_{B^\ast}},\ g_{B_s^\ast B_s^\ast\eta}=\frac{g_{B_sB_s^\ast\eta}}{m_{B_s^\ast}}\label{ff2}
\end{eqnarray}
with $f_\pi\simeq131$ MeV and $g_3\simeq0.569$, which is obtained by the measured decay width of $D^{\ast+}\to D^0\pi^+$ \cite{pdg2014}. All values of the coupling constants applied in our work are listed in Table \ref{tab:1}.
Other input parameters include the masses of the $\Upsilon(5S)$, the $\Upsilon(1^3D_1)$, the $\Upsilon(1^3D_2)$, th $\Upsilon(1^3D_3)$ and the $\eta$
which are $10.876$ GeV, $10.153$ GeV \cite{Eichten:1980mw}, $10.164$ GeV \cite{pdg2014}, $10.174$ GeV \cite{Eichten:1980mw} and $547.8$ MeV, respectively.

Finally, the general decay width of  $\Upsilon(5S)\to\Upsilon(1^3D_J)\eta$ reads
\begin{eqnarray}
\Gamma\left[\Upsilon(5S)\to\Upsilon(1^3D_J)\eta\right]=\frac{1}{3}\frac{1}{8\pi}\frac{|\bold{k}_2|}{m_{\Upsilon(5S)}^2}\overline{|\mathcal{A}|^2},\label{total}
\end{eqnarray}
where overlined bar indicates an average over the polarizations of the $\Upsilon(5S)$ in the initial state and sum over the polarizations of the $\Upsilon(1^3D_J)$ in the final state.
\begin{table}[htbp]
\caption{The values of the adopted coupling constants in our calculations. {Here, the values shown in the first row can be directly extracted by the partial decay widths of the $\Upsilon(5S)$ decays into $B^{(*)}_{(s)}\bar{B}^{(*)}_{(s)}$ in Ref. \cite{pdg2014}. The values listed in the second, the third, the fourth and fifth rows satisfy the relations constrained by heavy quark symmetry (see Eqs. (\ref{hhh1})-(\ref{hhh2}) and Eqs. (\ref{ff1})-(\ref{ff2})).}\label{tab:1}}
\footnotesize
\begin{center}
 \renewcommand{\arraystretch}{1.5}
 \tabcolsep=1.4pt
\begin{tabular}{c|cccccc}
\toprule[1pt]
$g$&            $BB$&              $BB^\ast$&            $B^\ast B^\ast$&            $B_sB_s$&            $B_sB_s^\ast$&            $B_s^\ast B_s^\ast$\\          \midrule[0.5pt]
$\Upsilon(5S)$& $1.2$&            $0.1$\ GeV$^{-1}$&       $2.1$&                     $1.0$&              $0.1$\ GeV$^{-1}$&         $6.7$\\
$\Upsilon(1^3D_1)$& $-427.2$&          $21.1$\ GeV$^{-1}$&       $43.1$&                     $-434.2$&              $21.5$\ GeV$^{-1}$&         $43.8$\\
$\Upsilon(1^3D_2)$& $\cdots$&      $407.2$&               $-13.4$\ GeV$^{-1}$&           $\cdots$&           $414.0$&                   $-13.6$\ GeV$^{-1}$\\
$\Upsilon(1^3D_3)$& $\cdots$&          $\cdots$&               $334.1$&               $\cdots$&        $\cdots$&                   $339.7$\\
$\eta$&         $\cdots$&         $18.8$&             $3.5$\ GeV$^{-1}$&              $\cdots$&              $-37.6$&                  $-6.9$\ GeV$^{-1}$\\
\bottomrule[1pt]
\end{tabular}
\end{center}
\end{table}


\section{Numerical results}\label{sec3}

\begin{figure}[hptb]
\begin{center}
\scalebox{1.0}{\includegraphics[width=\columnwidth]{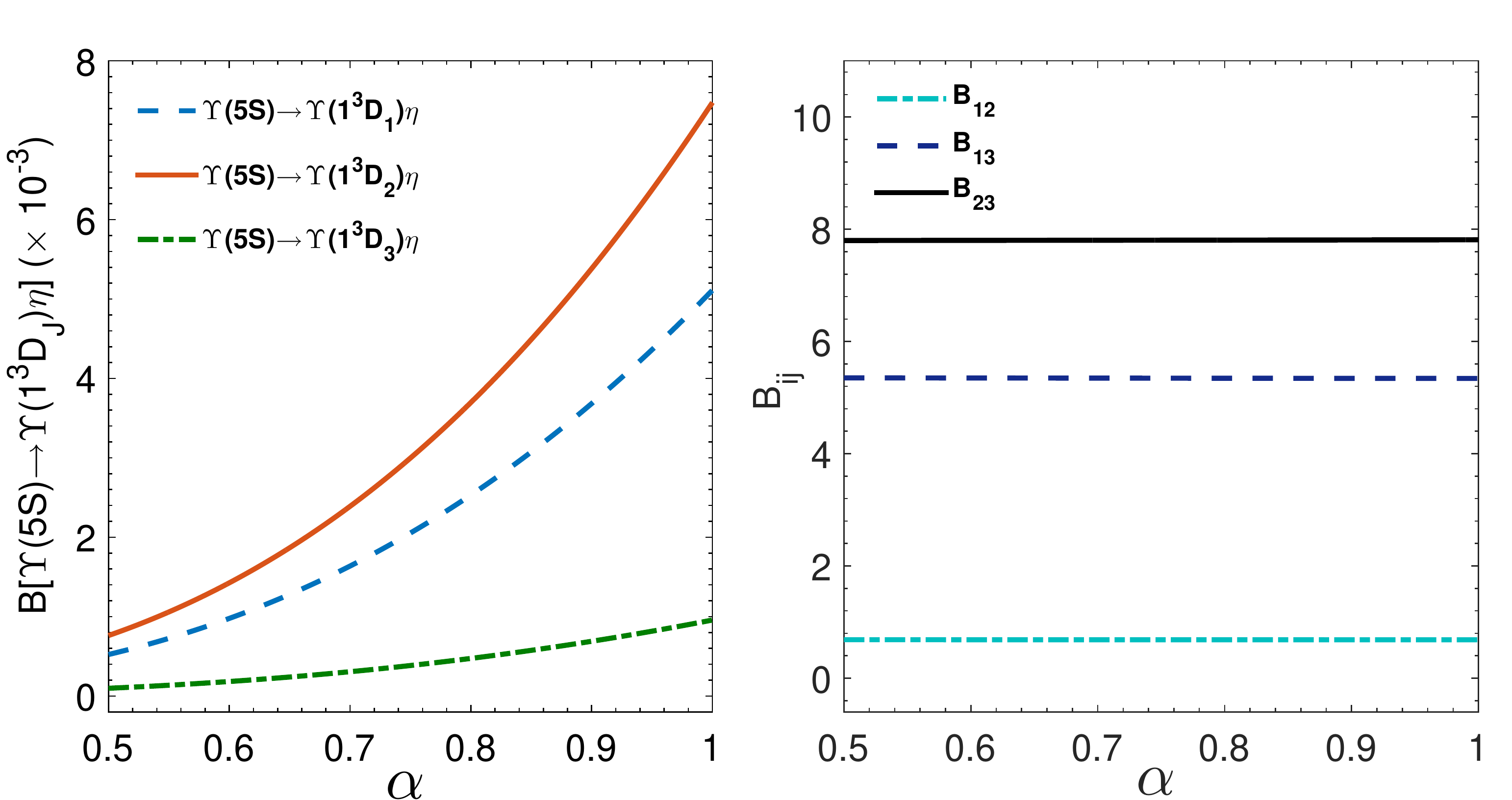}}
\end{center}
\caption{(color online). The dependence of the branching fractions of $\Upsilon(5S)\to\Upsilon(1^3D_{J})\eta$ on $\alpha$ (the left panel), and the dependence of the ratios $\mathcal{B}_{12}=\Gamma[\Upsilon(5S)\to\Upsilon(1^3D_1)\eta]/\Gamma[\Upsilon(5S)\to\Upsilon(1^3D_2)\eta]$, $\mathcal{B}_{23}=\Gamma[\Upsilon(5S)\to\Upsilon(1^3D_2)\eta]/\Gamma[\Upsilon(5S)\to\Upsilon(1^3D_3)\eta]$, and $\mathcal{B}_{13}=\Gamma[\Upsilon(5S)\to\Upsilon(1^3D_1)\eta]/\Gamma[\Upsilon(5S)\to\Upsilon(1^3D_3)\eta]$ on $\alpha$ (the right panel). }\label{BF}
\end{figure}

{Until now, we have fixed all coupling constants just shown in Sec. \ref{sec2}.
However, in our model there still exists a free parameter $\alpha$, which is introduced in Eq. (\ref{ff}) to parameterize the cutoff $\Lambda$.
Since the cutoff $\Lambda$ should be not deviated from the physical mass of the exchanged mesons, $\alpha$ is expected to be of order unity
just indicated in Ref. \cite{Cheng:2004ru}. Thus, in this work we take the range $0.5\leqslant\alpha\leqslant 1.0$ to present the numerical result of branching ratios for $\Upsilon(5S)\to\Upsilon(1^3D_J)\eta$, where the dependence of branching ratios on $\alpha$ are shown in the left panel of Fig. \ref{BF}.}

We predict
\begin{eqnarray}
\mathcal{B}[\Upsilon(5S)\to\Upsilon(1^3D_1)\eta]&=&(0.5\sim5.1)\times10^{-3},\nonumber\\
\mathcal{B}[\Upsilon(5S)\to\Upsilon(1^3D_2)\eta]&=&(0.7\sim7.5)\times10^{-3},\nonumber\\
\mathcal{B}[\Upsilon(5S)\to\Upsilon(1^3D_3)\eta]&=&(0.9\sim9.6)\times10^{-4},\nonumber
\end{eqnarray}
which are comparable with that of the observed $\Upsilon(5S)\to\Upsilon(nS)\pi^+\pi^-\ (n=1,2,3)$ \cite{Abe:2007tk}.
{In addition, we also notice that the Belle Collaboration reported a preliminary result of $\Upsilon(5S)\to\Upsilon(1D)\pi^+\pi^-$ \cite{Krokovny:2013}, where a production chain $\Upsilon(5S)\to\Upsilon(1D)\pi^+\pi^-\to\chi_b(1P)\gamma\pi^+\pi^-\to\Upsilon(1S)\gamma\gamma\pi^+\pi^-$ was measured, by which the branching ratio $\mathcal{B}[\Upsilon(5S)\to\Upsilon(1D)\pi^+\pi^-\to\Upsilon(1S)\gamma\gamma\pi^+\pi^-]=(2.0\pm0.4\pm0.3)\times10^{-4}$ was obtained. If combing this preliminary result with
data $\mathcal{B}[\Upsilon(1^3D_2)\to\chi_{b1}\gamma]=74.7\%$ \cite{Godfrey:2015dia} and $\mathcal{B}[\chi_{b1}\to\Upsilon(1S)\gamma]=(33.9\pm2.2)\%$ \cite{pdg2014}, we roughly get $\mathcal{B}[\Upsilon(5S)\to\Upsilon(1D)\pi^+\pi^-]=(0.5\sim1)\times10^{-3}$.
If checking decay behaviors of $\Upsilon(5S)\to\Upsilon(nS)\pi^+\pi^-(\eta)~(n=1,2)$ \cite{pdg2014,Krokovny:2013}, we may find the dipion and $\eta$ transitions of $\Upsilon(5S)$ are of the same order of magnitude.
These predicted branching ratios of
$\Upsilon(5S)\to\Upsilon(1^3D_J)\eta$ in this work are reasonable if comparing it with $\Upsilon(5S)\to\Upsilon(1D)\pi^+\pi^-$.
Thus, it is obvious that there may exist anomalous $\Upsilon(5S)\to \Upsilon(1^3D_J)\eta$ transitions, and we hope future experiment like Belle and BelleII can test our predictions.

In addition, we also obtain several typical ratios, {which are independent of the coupling constant $g_2$, and can reflect the relative magnitude of these predicted branching ratios of $\Upsilon(5S)\to\Upsilon(1^3D_{J})\eta$. Just as shown in the right panel of Fig. \ref{BF}, these obtained ratios are weakly dependent on $\alpha$. Thus, we list their values below
\begin{eqnarray}
\mathcal{B}_{12}&=&{{\Gamma[\Upsilon(5S)\to\Upsilon(1^3D_1)\eta]}\over{\Gamma[\Upsilon(5S)\to\Upsilon(1^3D_2)\eta]}} =0.68,\nonumber\\
\mathcal{B}_{13}&=&{\Gamma[\Upsilon(5S)\to\Upsilon(1^3D_1)\eta]\over\Gamma[\Upsilon(5S)\to\Upsilon(1^3D_3)\eta]}=5.34,\nonumber\\
\mathcal{B}_{23}&=&{\Gamma[\Upsilon(5S)\to\Upsilon(1^3D_2)\eta]\over\Gamma[\Upsilon(5S)\to\Upsilon(1^3D_3)\eta]}=7.79.\nonumber
\end{eqnarray}

Among these $\Upsilon(5S)\to\Upsilon(1^3D_{J})\eta$ decays, $\Upsilon(5S)\to\Upsilon(1^3D_{1})\eta$ and $\Upsilon(5S)\to\Upsilon(1^3D_{2})\eta$ have the same order of magnitude, while $\Upsilon(5S)\to\Upsilon(1^3D_{3})\eta$ is about one order of magnitude smaller than these two decays.

\section{Discussion and Conclusion}\label{sec4}

With the experimental progress on the hidden-bottom decays of the $\Upsilon(5S)$ \cite{Abe:2007tk,He:2014sqj,Belle:2011aa,Adachi:2011ji}, the importance of hadronic loop mechanism to the $\Upsilon(5S)$ decays has been realized by theorist \cite{Meng:2007tk,Meng:2008dd,Chen:2011qx,Chen:2011zv,Chen:2011pv,Meng:2008bq,Chen:2014ccr}. When studying the processes relevant to higher bottomonia, the coupled-channel effect cannot be ignored since more hadronic channels are open.

In this work, the hadronic loop mechanism, which is an equivalent description of coupled-channel effect,
was applied to the study of the $\Upsilon(5S)\to\Upsilon(1^3D_J)\eta\ (J=1,2,3)$ transitions. Our calculation shows that  anomalous $\Upsilon(5S)\to\Upsilon(1^3D_J)\eta\ (J=1,2,3)$ transitions may exist, which can be tested by future experiment like Belle and forthcoming BelleII. Associated with former studies of the hadronic decays of the $\Upsilon(5S)$ \cite{Meng:2007tk,Meng:2008dd,Simonov:2008qy,Chen:2011qx,Chen:2011zv,Chen:2011pv,Meng:2008bq,Chen:2014ccr}, the present work again proves that the hadronic loop mechanism as one of non-perturbative QCD effects indeed plays important role in understanding anomalous hidden-bottom decays of the $\Upsilon(5S)$.

We need to specify that how to experimentally identify D-wave bottomonia $\Upsilon(1^3D_J)$ ($J=1,2,3$) will be the first step of exploring the $\Upsilon(5S)\to\Upsilon(1^3D_J)\eta$ transitions. When analyzing the $\Upsilon(5S)\to\Upsilon(1^3D_J)\eta$ decays, it is easy way to reconstrute the $\eta$ by the $\eta\to \gamma\gamma$ process. And then, the $M_{miss}(\eta)$ distribution can be obtained. By analyzing $M_{miss}(\eta)$ distribution, the signals of $\Upsilon(1^3D_J)$ can be found. This approach was adopted in studying the $\Upsilon(4S)\to\eta h_b(1P)$ decay \cite{Tamponi:2015xzb}.
In Table \ref{tab:2}, we list some theoretical values of the masses of D-wave bottomonia $\Upsilon(1^3D_J)$, by which we notice that there exist small mass splittings between $D$-wave bottomonia $\Upsilon(1^3D_J)$. Thus, there should exist some difficulties to distinguish different $\Upsilon(1^3D_J)$ states by the obtained $M_{miss}(\eta)$ distribution
of $\Upsilon(5S)\to\Upsilon(1^3D_J)\eta$ since the $\Upsilon(1^3D_J)$ states are close to each other, which is the challenge to experimental analysis of $\Upsilon(5S)\to\Upsilon(1^3D_J)\eta$.

In summary, searching for the predicted $\Upsilon(5S)\to\Upsilon(1^3D_J)\eta$ transitions will be a potential and intriguing research issue to future experiment, which will make the study of the hadronic transitions of higher bottomonia become more abundant. What is more important is that the search for the $\Upsilon(5S)\to\Upsilon(1^3D_J)\eta$ can further test the importance of hadronic contribution to the $\Upsilon(5S)$ decays.
We are waiting for the experimental progress on this issue.
\begin{table}[htbp]
\caption{The calculated masses of bottomonia $\Upsilon(1^3D_J)$ $(J=1,2,3)$ from different theoretical models.\label{tab:2}}
\begin{center}
 \renewcommand{\arraystretch}{1.5}
 \tabcolsep=1.4pt
\begin{tabular}{c|cccccc}
\toprule[1pt]
State&            Ref. \cite{Eichten:1980mw}&      Ref. \cite{Ebert:2002pp}&    Ref. \cite{Godfrey:2015dia}&    Ref. \cite{Gupta:1982kp}&       Ref. \cite{Moxhay:1983vu}&      Ref. \cite{Kwong:1988ae}   \\          \midrule[0.5pt]
$\Upsilon(1^3D_1)$& $10.153$&       $10.153$&         $10.138$&        $10.155$&       $10.151$&          $10.150$\\
$\Upsilon(1^3D_2)$& $10.163$&       $10.158$&         $10.147$&       $10.162$&        $10.161$&          $10.156$\\
$\Upsilon(1^3D_3)$& $10.174$&       $10.162$&          $10.155$&      $10.167$&        $10.168$&          $10.160$\\
\bottomrule[1pt]
\end{tabular}
\end{center}
\end{table}

\section*{Appendix A: The constructions of polarization tensors}\label{appendixA}
The polarization tensor of high spin state, which has angular momentum $j$ and helicity $\lambda$, can be constructed with polarization vectors and Clebsch-Gordan coefficients. In general, there exists a relation \cite{Zhu:1999ur},
\begin{eqnarray}\label{polsum}
\epsilon_{\mu_1\cdots\mu_n}(p,\lambda)&=&\sum_{\lambda_{n-1},\lambda_n}\langle n-1,\lambda_{n-1};n,\lambda_n\rangle\nonumber\\
&&\times\epsilon_{\mu_1\mu_2\cdots\mu_{n-1}}(p,\lambda_{n-1})\epsilon_{\mu_n}(p,\lambda_n),
\end{eqnarray}
where $\lambda=-n,-n+1,\cdots, n$, and the polarization tensor must satisfy the following conditions,
\begin{eqnarray}
\text{Transverse}: p^{\mu_i}\epsilon_{\mu_1\cdots\mu_i\cdots\mu_n}(p,\lambda)&=&0,\nonumber\\
\text{Symmetric}:\epsilon_{\mu_1\cdots\mu_i\cdots\mu_j\cdots\mu_n}(p,\lambda)&=&\epsilon_{\mu_1\cdots\mu_j\cdots\mu_i\cdots\mu_n}(p,\lambda),\nonumber\\
\text{Traceless}:g^{\mu_i\mu_j}\epsilon_{\mu_1\cdots\mu_i\cdots\mu_j\cdots\mu_n}(p,\lambda)&=&0,\nonumber\\
\epsilon^\ast_{\mu_1\cdots\mu_n}(p,\lambda)\epsilon^{\mu_1\cdots\mu_n}(p,\lambda^\prime)&=&(-1)^n\delta_{\lambda\lambda^\prime},\nonumber\\
\epsilon^\ast_{\mu_1\cdots\mu_n}(p,\lambda)&=&(-1)^\lambda\epsilon^{\mu_1\cdots\mu_n}(p,-\lambda).\nonumber
\end{eqnarray}

The detailed form of the polarization vectors of the $j=1$ state with different helicity is
\begin{eqnarray}\label{polvec}
\epsilon^\mu(\pm1)=(0,\mp1,-i,0)/\sqrt{2},\ \ \ \  \epsilon^\mu(0)=(p_f^3,0,0,p_f^0)/m_f.
\end{eqnarray}
According to Eq. (\ref{polsum}), we further write out the polarization tensors, i.e.,
\begin{subequations}
\begin{eqnarray}
\epsilon^{\mu\nu}(\pm2)&=&\epsilon^{\mu}(\pm1)\epsilon^{\nu}(\pm1),\label{poltensor2}\\
\epsilon^{\mu\nu}(\pm1)&=&\sqrt{\frac{1}{2}}\left[\epsilon^\mu(\pm1)\epsilon^\nu(0)+\epsilon^\mu(0)\epsilon^\nu(\pm1)\right],\label{poltensor1}\\
\epsilon^{\mu\nu}(0)&=&\sqrt{\frac{1}{6}}\left[\epsilon^\mu(+1)\epsilon^\nu(-1)+\epsilon^\mu(-1)\epsilon^\nu(+1)\right]\nonumber\\
&&+\sqrt{\frac{2}{3}}\epsilon^\mu(0)\epsilon^\nu(0)\label{poltensor0}
\end{eqnarray}
\end{subequations}
for  of $j=2$ state, and
\begin{subequations}
\begin{eqnarray}
\epsilon^{\alpha\beta\gamma}(\pm3)&=&\epsilon^{\alpha\beta}(\pm2)\epsilon^\gamma(\pm1),\\
\epsilon^{\alpha\beta\gamma}(\pm2)&=&\frac{1}{\sqrt{3}}\epsilon^{\alpha\beta}(\pm2)\epsilon^\gamma(0)+\sqrt{\frac{2}{3}}\epsilon^{\alpha\beta}(\pm1)\epsilon^\gamma(\pm1),\\
\epsilon^{\alpha\beta\gamma}(\pm1)&=&\frac{1}{\sqrt{15}}\epsilon^{\alpha\beta}(\pm2)\epsilon^\gamma(\mp1)+2\sqrt{\frac{2}{15}}\epsilon^{\alpha\beta}(\pm1)\epsilon^\gamma(0)\nonumber\\
&&+\sqrt{\frac{2}{5}}\epsilon^{\alpha\beta}(0)\epsilon^\gamma(\pm1),\\
\epsilon^{\alpha\beta\gamma}(0)&=&\frac{1}{\sqrt{5}}\epsilon^{\alpha\beta}(+1)\epsilon^\gamma(-1)+\sqrt{\frac{3}{5}}\epsilon^{\alpha\beta}(0)\epsilon^\gamma(0)\nonumber\\
&&+\frac{1}{\sqrt{5}}\epsilon^{\alpha\beta}(-1)\epsilon^\gamma(+1)
\end{eqnarray}
\end{subequations}
for $j=3$ state. Here, the expressions of the vector $\epsilon^\gamma(\lambda)$ and the tensor $\epsilon^{\alpha\beta}(\lambda)$ have been given in Eq. (\ref{polvec}) and Eqs. (\ref{poltensor2})-(\ref{poltensor0}).


\section{Acknowledgments}

We would like to thank Estia Eichten and Susana Coito for useful discussions. This project is supported by the National Natural Science Foundation of China under Grants No.~11222547, No.~11175073, No.~11375240, and No.~11035006, and by Chinese Academy of Sciences under the founding Y104160YQ0 and the agreement No.~2015-BH-02. XL is also supported by the National Program for Support of Top-notch Young Professionals.

\end{document}